\documentclass[a4paper]{jpconf}
\usepackage{graphicx}

\usepackage{citesort}

\begin{document}

\title{Imaging ultrafast electronic motion by x-ray scattering}

\author{Gopal Dixit$^{1, 2}$ and Robin Santra$^{1, 2, 3}$}

\address{$^{1}$ Center for Free-Electron Laser Science, DESY, Notkestrasse 85, 22607 Hamburg, Germany}
\address{$^{2}$ The Hamburg Centre for Ultrafast Imaging, Luruper Chaussee 149, 22761 Hamburg, Germany}
\address{$^{3}$ Department of Physics, University of Hamburg, 20355 Hamburg, Germany}

\ead{gopal.dixit@cfel.de; robin.santra@cfel.de}

\begin{abstract}
Time-resolved ultrafast x-ray scattering is an emerging approach
to probe the temporally evolving electronic charge distribution
in real-space and in real-time. In this contribution,
time-resolved ultrafast x-ray scattering from an electronic wave 
packet is presented. It is shown that the spatial and temporal
correlations are imprinted in the scattering patterns, obtained by
ultrafast x-ray scattering from an electronic wave packet, which
deviate drastically from the notion that the instantaneous
electronic density is the key quantity being probed. Furthermore,
a detailed analysis of ultrafast x-ray scattering from a sample
containing a mixture of non-stationary and stationary electrons
along with the role of scattering interference between a
non-stationary and several stationary electrons to the total
scattering signal is discussed.
\end{abstract}

\section{Introduction}
Owing to the availability of laser pulses on the sub-fs (1 fs =
10$^{-15}$ s) time scale~\cite{hentschel,goulielmakis1}, much
progress has been accomplished in imaging the electronic wave 
packet dynamics in real-time with table-top experiments in recent
years~\cite{haessler,tzallas,hockett,goulielmakis}. One of the
important ambitions of the emerging field of time-resolved imaging
is to image the motion of an electron not only in real-time but
also in real-space. Imaging the electronic motion with atomic-scale 
spatial and temporal resolution will help to fully
understand the functionality and dynamic behavior of atoms,
complex molecules and solids, for example several ubiquitous
ultrafast phenomena such as bond formation and breaking, photoinduced
exciton dynamics, and conformational changes and charge
migration~\cite{breidbach2003,kuleff2005,remacle}.

\mbox{X-ray} crystallography is a well-established method in
various areas of science to obtain real-space, atomic-scale
structural information of complex materials, ranging from
molecular systems~\cite{Ihee} to proteins~\cite{Chapman}. With the
tremendous advancement in technology for producing ultrashort,
ultraintense and tunable x-ray pulses from free-electron lasers
(XFELs), laser plasmas and high-harmonic
generation~\cite{emma2,ishikawa2012,popmintchev2012}, it seems
feasible to obtain information about ultrafast dynamics of
electrons. In recent years, using the unprecedented properties of
x-rays from XFELs, several experiments have been carried out for
systems ranging from atoms~\cite{young2010,rohringer2012},
molecules~\cite{hoener2010,berrah2011}, to complex
biomolecules~\cite{Chapman,seibert}. In order to image the electronic motion
in real-space and in real-time, pump-probe experiment
seems the most convenient approach. In such experiment, a pump pulse
induces the dynamics and then subsequently a probe pulse
interrogates such induced dynamics. One can perform scattering of
ultrashort \mbox{x-ray} pulses from the temporally evolving
electronic charge distribution. A series of scattering patterns
obtained by varying the pump-probe time-delay serve to image the
electronic motion with atomic-scale spatial and temporal
resolution. However, at this juncture a fundamental question needs
to be addressed: How does an ultrashort x-ray pulse interact and
scatter from a temporally evolving quantum system?

\section{Theory}
In the stationary case, the differential scattering probability
(DSP), which is a key quantity in x-ray scattering, is related to 
the Fourier transform of the electron density $\rho_{e}(\mathbf{x})$ as 
\begin{equation}\label{eq01}
\frac{dP}{d\Omega} = \frac{dP_{e}}{d\Omega}
\left|\int d^{3}x\; \rho_{e}(\mathbf{x}) ~ e^{i\mathbf{Q \cdot
x}} \right|^{2}.
\end{equation}
Here, ${dP_{e}}/{d\Omega}$ is the DSP for a free electron and
$\mathbf{Q}$ is the photon momentum transfer. 
In the recent past, the idea of obvious extension of the x-ray
scattering from static to time domain was proposed by
assigning an additional degree of freedom to the electron density in Eq.~(\ref{eq01}),
i.e., replacing $\rho_{e}(\mathbf{x})$ by $\rho_{e}(\mathbf{x},
t)$~\cite{Krausz,jurek}. Let us first employ the semiclassical
theory of light-matter interaction to describe the scattering of
ultrashort x-ray pulse from an electronic wave packet. In the
semiclassical theory, matter is treated quantum mechanically,
whereas x-ray is treated classically. According to the
semiclassical theory, the DSP can be
expressed as~\cite{dixit2012}
\begin{equation}\label{eq1}
\frac{dP}{d\Omega} = \frac{dP_{e}}{d\Omega}
\left|\int d^{3}x\; \rho_{e}(\mathbf{x}, t) ~ e^{i\mathbf{Q \cdot
x}} \right|^{2}.
\end{equation}
Here, we have assumed that the x-ray pulse duration is shorter than the
dynamical time scale of the electronic wave packet.  According to
Eq.~(\ref{eq1}), the measured scattering pattern provides access
to the instantaneous electron density, $\rho_{e}(\mathbf{x}, t)$,
as a function of the pump-probe delay time $t$.
Equation~(\ref{eq1}) imposes the constraint that the state of the
electronic wave packet remains unchanged before and after the
scattering process. But one needs an ultrashort \mbox{x-ray} pulse
to probe electronic motion on an ultrafast time scale, which has
an unavoidable bandwidth. Due to the finite bandwidth of an
ultrashort \mbox{x-ray} pulse, it is fundamentally impossible to
detect whether or not a transition between the eigenstates spanned
by the wave packet, or transitions to other states closer in
energy than the pulse bandwidth, has taken place. As a result, it
is energetically impossible to detect inelasticity within the
bandwidth. Therefore, unavoidable inelasticity leads to
contributions from distinguishable final states (eigenstates)
within the bandwidth, and these contributions must be summed
incoherently.

In order to overcome the constraint imposed by Eq.~(\ref{eq1}), we now employ a
consistent full quantum theory of light-matter interaction.
In the fully quantum mechanical theory, matter and x-ray
both are treated quantum mechanically and first-order
time-dependent perturbation theory is employed for the interaction
between matter and x-rays. The resulting expression for the DSP
from a coherent, Gaussian \mbox{x-ray} pulse is~\cite{dixit2012}
\begin{eqnarray}\label{eq2}
\frac{dP}{d\Omega} & = &  \frac{dP_{e}}{d\Omega} \int_{0}^{\infty}
d\omega_{\mathbf{k}_{s}} ~ W_{\Delta E}
({\omega_{\mathbf{k}_{s}}}) ~
\frac{\omega_{\mathbf{k}_{s}}}{\omega_{\mathbf{k}_{in}}} ~
\int_{-\infty}^{\infty} \frac{d\tau}{2\pi}
~e^{-(\frac{2\ln{2}\;\tau^{2}}{\tau_{l}^{2}})}~ e^{-
i(\omega_{\mathbf{k}_{s}}-
\omega_{\mathbf{k}_{in}}) \tau }\nonumber \\
& & \times \int d^{3}x \int d^{3}x^{\prime} ~ \left \langle \Psi
\left( t +\frac{\tau}{2} \right) \Biggl| ~
\hat{n}\left(\mathbf{x}^{\prime} \right)~
e^{-i\hat{H}\tau}~\hat{n}\left(\mathbf{x}\right)~ \Biggr| \Psi
\left( t-\frac{\tau}{2}\right) \right \rangle e^{i{\mathbf{Q}}
\cdot (\mathbf{x}-\mathbf{x^\prime})}.
\end{eqnarray}
Here, $\omega_{\mathbf{k}_{in}}$ and $\omega_{\mathbf{k}_{s}}$ refer to the energy of the incident and scattered
photon, respectively, and $\tau_{l}$ is the pulse duration.
$\hat{H}$ represents the electronic
Hamiltonian, $\hat{n}(\mathbf{x})$ is the electron density operator and
$W_{\Delta E}({\omega_{\mathbf{k}_{s}}})$ is a
spectral window function centered at $\omega_{\mathbf{k}_{in}}$
with a width $\Delta E$, which decides the range of energies of the scattered photons accepted by
the detector.

\section{Results and Discussions}
It is evident from Eqs.~(\ref{eq1}) and ~(\ref{eq2}) that the
semiclassical and full quantum theory provide completely different
results about the electronic wave packet motion. To visually
demonstrate the difference between both the equations, we
calculate the scattering patterns corresponding to an electronic
wave packet.

\subsection{Time-resolved x-ray scattering from an electronic wave packet in atomic hydrogen}
\begin{figure*}
\begin{center}
\includegraphics[width=15cm]{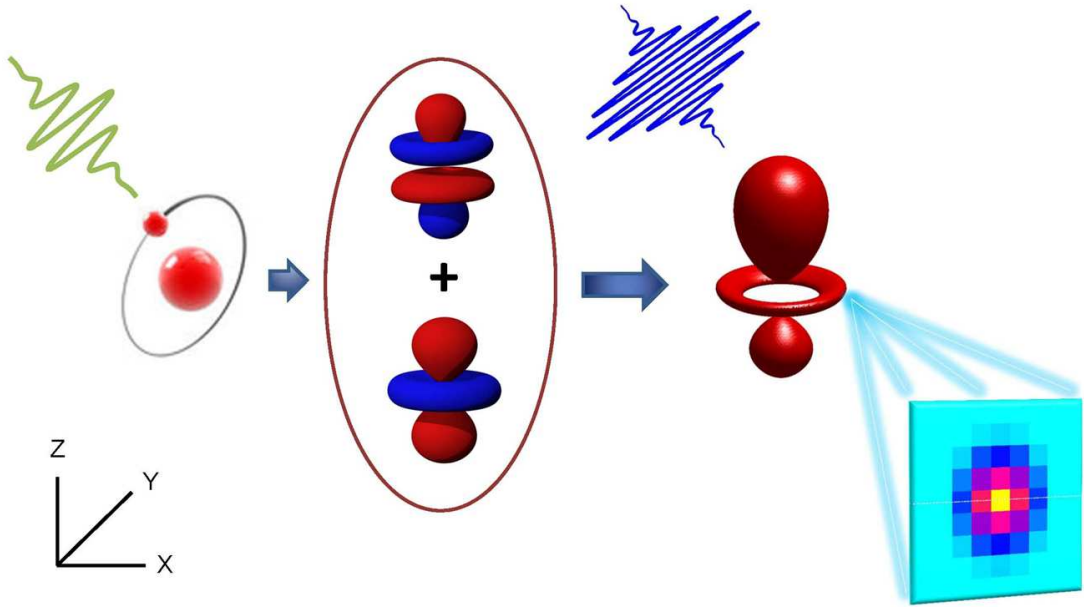}
\end{center}
\caption{Schematic of time-resolved ultrafast x-ray scattering for imaging an electronic
wave packet motion with atomic-scale spatial and temporal resolution. 
Adapted from Dixit {\it et al.}~\cite{dixit2012}.} \label{fig1}
\end{figure*}

A schematic scenario for imaging an electronic wave packet via
ultrafast x-ray scattering is shown in Fig~\ref{fig1}. A pump
pulse with broad energy bandwidth prepares a coherent
superposition  with equal population of the 3d and 4f eigenstates
of atomic hydrogen with projection of orbital angular momentum
equal to zero. The electronic charge distribution undergoes
periodic oscillation with oscillation period T = 6.25 fs. The
dynamically evolving charge distribution is imaged via ultrafast
x-ray scattering.

\begin{figure*}
\begin{center}
\includegraphics[width=15cm]{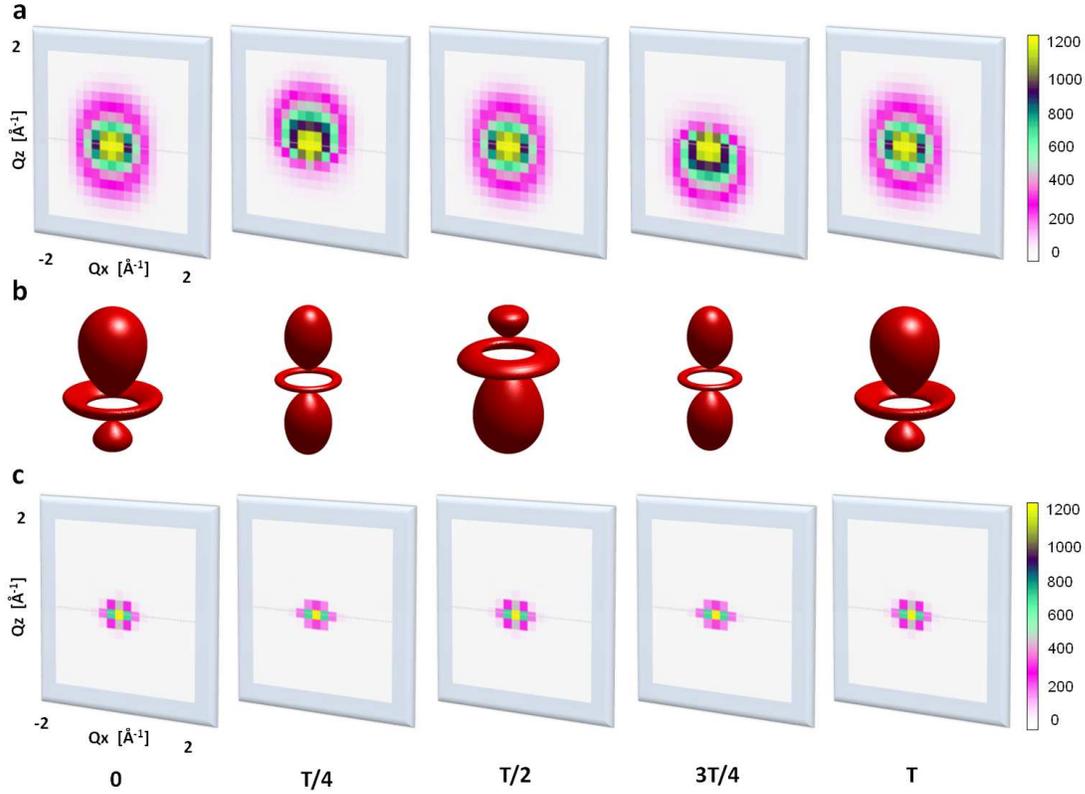}
\end{center}
\caption{Time-resolved scattering pattern in the $Q{_x}$ - $Q{_z}$
plane ($Q{_y}$ = 0) and electronic charge distribution of the
wave packet. (a) Scattering pattern obtained using Eq.~(\ref{eq2}),
(b) isosurface of the electronic charge distribution and (c)
scattering pattern obtained using Eq.~(\ref{eq1}), at the
pump-probe delay times at times 0, T/4, T/2, 3T/4, and T, where T
= 6.25 fs is the oscillation period. The patterns
are shown in units of the Thomson scattering cross section in both
cases. Adapted from Dixit {\it et al.}~\cite{dixit2012}.} \label{fig2}
\end{figure*}

As a function of the delay time at times 0, T/4, T/2, 3T/4, and T,
the scattering pattern in the $Q_x$ - $Q_z$ plane ($Q_y$ = 0) of
the electronic wave packet is depicted in Fig.~\ref{fig2}. The
patterns shown in Fig.~\ref{fig2}a and Fig.~\ref{fig2}c are
computed using Eqs.~(\ref{eq2}) and ~(\ref{eq1}), respectively.
The isosurface of the electronic charge distribution shown in
Fig.~\ref{fig2}b encloses $\sim 26 \%$ of the total probability
and has length 14--17 \AA~ along the $z$-axis and 7.5--9 \AA~
along the $x$ and $y$-axes. The wave packet is exposed to a 1-fs
\mbox{x-ray} pulse with 4~keV photons and we assume a Gaussian
photon energy detection window of width $\Delta E=$~1~eV for the
detector. Therefore, the contributions from all electronic
transitions within 1 eV energy range are included incoherently for
the scattering patterns shown in Fig.~\ref{fig2}a. The patterns are
calculated for ${Q}_{\mathrm{max}} = 2$ \AA$^{-1}$ corresponding
to a 3.14 \AA~spatial resolution and to a detection angle of
scattered photons of up to 60$^{\circ}$.

The patterns in Fig.~\ref{fig2}a undergo spatial oscillation along
$Q_z$ and mimic the wave packet motion along $z$ in real space as
is evident from Fig.~\ref{fig2}b. However, a careful observation
reveals the striking feature that the patterns are asymmetric when the 
corresponding charge distributions are symmetric and vice versa,
which can be explained as follows: The electron clouds move in
opposite directions at delay times T/4 and 3T/4, as may be seen in
Fig.~\ref{fig2}b, while the charge distributions are identical at
the two times.  At time T/4, the flow of the electron cloud is
downwards, whereas at time 3T/4 the flow is upwards. This is
reflected by their corresponding patterns. Moreover, the patterns
in Fig.~\ref{fig2}a are not centrosymmetric at all times, 
i.e., they are not equal for $\mathbf{Q}$ and $-\mathbf{Q}$. This is in
contrast with the centrosymmetric patterns expected from
Eq.~(\ref{eq1}) as a consequence of Friedel's law~\cite{Nielsen},
and shown in Fig.~\ref{fig2}c. This counterintuitive nature of the
scattering patterns arises from the fact that they are not simply
related to the Fourier transform of the instantaneous electronic
density, but they are related to the electronic wave packet through
Eq.~(\ref{eq2}). Clearly, the patterns in Fig.\ref{fig2}c contain
less information than the patterns shown in Fig.~\ref{fig2}a and
they miss important phase information. Therefore, the patterns
calculated within the full quantum theory capture the dynamics of
the {\em momentum distribution} of the wave packet. As a
consequence, the apparent motions of the charge distributions and
of the scattering patterns are phase shifted by 90$^{\circ}$. On the
other hand, the patterns shown in Fig.~\ref{fig2}c do not change
significantly as a function of the delay time.

\begin{figure*}
\begin{center}
\includegraphics[width=15cm]{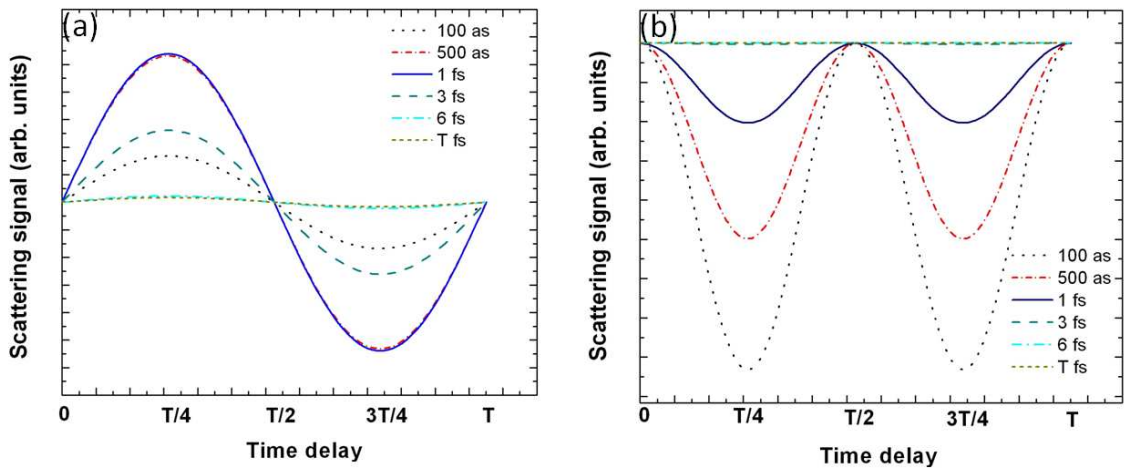}
\end{center}
\caption{
Effect of x-ray pulse duration
on the scattering pattern as a function of the delay time. The scattering signal
at an individual pixel of the scattering pattern obtained (a) using Eq.~(\ref{eq2}), and (b) using Eq.~(\ref{eq1})
for different pulse durations. Adapted from Dixit {\it et al.}~\cite{dixit2012}.} \label{fig3}
\end{figure*}

Not only the scattering patterns obtained using Eqs.~(\ref{eq1})
and ~(\ref{eq2}) are completely different, but the effect of the
x-ray pulse duration is also different. For different pulse
durations, the scattering signal at an individual pixel of the
scattering pattern is shown in Fig.~\ref{fig3}. The scattering
signal calculated with Eq.~(\ref{eq2})  has the same periodicity
as the corresponding wave packet dynamics. In contrast, the
scattering signal calculated with Eq.~(\ref{eq1}) has the wrong
periodicity. The effect of pulse duration for long pulses
approaching the period of the dynamics being probed is similar for
both the theories as shown in Fig.~\ref{fig3}. In this case, the
scattering signal averages out and there is no dynamical
information. There is no optimum pulse duration for the signal
computed with Eq.~(\ref{eq1}) in going to short pulses. The
contrast as a function of time increases monotonically and for
short enough pulses becomes almost constant. In contrast, the
signal computed with Eq.~(\ref{eq2}) has an optimum contrast as a
function of time for a pulse duration around 1-fs. For shorter
pulses, the signal flattens again. The reason can be found in the
matrix element in the second line of Eq.~(\ref{eq2}). This matrix
element becomes independent of $\mathbf{Q}$ for $\tau \rightarrow
0$, intuitively because the electron has no time to change its
position. This is the only contribution summed over for a pulse of
length $\tau_l \rightarrow 0$, and consequently, the signal
becomes independent of the electronic dynamics.

\subsection{Role of electron-electron interference}
Time-resolved ultrafast x-ray scattering from a one-electron
system is not a realistic situation. In reality, when a broad
energy bandwidth pump pulse illuminates an $N$-electron system, one or
few electrons form an electronic wave packet and other electrons
remain stationary. In such scenario, when x-rays scatter from a
mixture of stationary and non-stationary electrons, it is not
possible to figure out whether the scattering occurs from the
non-stationary electrons or from the stationary electrons and how
the interference between non-stationary and stationary electrons
contributes to the total scattering signal. Hence, it is crucial
to analyze different types of contributions in the scattering
process. In an $N$-electron system, different scattering
contributions to the total scattering signal within both the
theories  have been rigorously analyzed by us in
Ref.~\cite{dixit2013jcp}. More specifically, the motion of a
one-electron wave packet in the presence of $N$ stationary
electrons was investigated. In this case, $N$ stationary electrons
serve as reference scatterers in the total scattering signal.

It has been shown that the total signal can be  factored into three main
parts: first from stationary electrons, second from non-stationary
electrons and third from the interference between non-stationary
and stationary electrons. Also, it has been demonstrated that the scattering
contributions from the stationary electrons to the signal are
identical, whereas scattering contributions from the
non-stationary electron are completely different in both the
theories~\cite{dixit2013jcp}. Moreover, the important contribution
from the scattering interference is also different within both the
theories. The scattering interference within the full quantum
theory entirely depends on the energy resolution of the detector
and the x-ray pulse duration. In case of extremely short pulses or
negligible energy resolution, full quantum theory provides
identical contributions for the scattering interference as one
obtains in the semiclassical theory. In contrast to that, if the
x-ray pulse is not very short in comparison to the dynamical
time scale of the motion and, if the energy resolution of the
detector is sufficiently high, the scattering interference within
the full quantum theory does not provide identical result to the
one obtained in the semiclassical theory~\cite{dixit2013jcp}.

After analyzing the role of scattering interference between
non-stationary and stationary electrons, time-resolved ultrafast
x-ray scattering from an electronic wave packet in helium was
investigated~\cite{dixit2013jcp}. In helium, a pump pulse excites
one of the electrons from the ground state and forms a coherent
superposition with equal population of the 1s3d and 1s4f
configurations with the projection of orbital angular momentum
being equal to zero. 
The energy difference between the 1s3d and
1s4f configurations is 0.66 eV, and 
the ground
state and 1s3d configurations is 23.07 eV. Also, the energy difference
between non-stationary and stationary electrons is around 23 eV
and they are energetically distinguishable. Therefore, any excitation
from the stationary electron can be easily filtered out in the
energy-resolved scattering process and hence not considered in the present case. 
The scattering patterns have been computed for the non-stationary electron in the presence of a
stationary electron using both the theories. The time-dependent
interference between the stationary and non-stationary electrons
within the semiclassical theory is zero, and it is quite small in
comparison to the total scattering signal in the full quantum
theory. However, the time-independent interference between the
stationary and non-stationary electrons contributes identically to
the total signal in both the theories. The patterns are dominated
by the scattering contribution from the time-independent
interference within the semiclassical theory, whereas the patterns
are dominated by the scattering contributions from the
non-stationary electron due to Compton scattering within the full
quantum theory.

\subsection{An alternative way to image electronic motion}
After analyzing the two different cases for time-resolved x-ray
scattering, it has been shown that the scattering patterns encode the
spatial-temporal density-density correlations, not the
instantaneous electronic density. 
Hence, it is not easy to obtain direct information about the dynamical structural changes
during the complex processes from these correlations. 
Whereas, if one can retrieve the instantaneous electronic density, 
it will provide more direct insight about these processes.
Therefore, one may raise
question about the possibility to image instantaneous electronic
density of the wave packet in real-space and in real-time using
x-ray. To answer this question, a possible way
to image the instantaneous electron density via ultrafast
\mbox{x-ray} {\em phase contrast imaging} has been proposed by
us~\cite{dixit2013prl}. 
Ultrafast \mbox{x-ray} phase contrast
imaging is based on the interference between 
two pathways -- incident and scattered -- of an x-ray photon.
Therefore, this method does not suffer from the
problem of inelastic scattering processes within the finite
bandwidth of the pulse. The key quantity in this method is the
Laplacian of the projected instantaneous electronic density of the
wave packet.  Our
proposed method offers a potential to image not only instantaneous
snapshots of non-stationary electron dynamics, but also the
Laplacian of these snapshots which reveals the internal structures
of the wave packet through local variations in the instantaneous
electronic density and provides information about the complex
bonding and topology of the charge distributions in the
systems~\cite{dixit2013prl}.

\section{Conclusion}
The illustrative example used as a proof of principle is
physically simple, but lies in the time and energy range of
interest corresponding to the dynamics of valence electrons in
more complex molecular and biological systems. Interestingly, the
scattering patterns provide an unusually visual manifestation of
the quantum nature of light. This quantum nature becomes central
only for non-stationary electronic states and has profound
consequences for time-resolved imaging. Our present findings on
time-resolved ultrafast x-ray scattering will find several
important applications for exploring ultrafast dynamics in nature.

\section{Acknowledgments}
We thank Oriol Vendrell and  Jan Malte Slowik for fruitful collaboration on the topics presented here. 

\section*{References}
%\bibliography{imaging}
\providecommand{\newblock}{}

\end{document}